\documentclass[aps,prl,twocolumn,showpacs,superscriptaddress,groupedaddress]{revtex4}  
\usepackage{graphicx}  
\usepackage{dcolumn}   
\usepackage{bm}        
\usepackage{amssymb}   

\def\icg{Institute of Cosmology and Gravitation, University of Portsmouth, Burnaby Rd, Portsmouth, PO1 3FX, UK}

\newcommand{\be}{\begin{equation}}
\newcommand{\ee}{\end{equation}}
\newcommand{\symbl}{{{n}_g}}
\newcommand{\citepeg}[1]{\citep[e.g.][]{#1}}
\newcommand{\comment}[1]{}
\newcommand{\mnras}{MNRAS}

\begin{document}

\title{Testing the speed of gravitational waves over cosmological distances with strong gravitational lensing}
\author{Thomas E. Collett \& David Bacon}
\affiliation{\icg}
\email{thomas.collett@port.ac.uk}
\begin{abstract}

Probing the relative speeds of gravitational waves and light acts as an important test of General Relativity and alternative theories of gravity. Measuring the arrival time of gravitational waves and electromagnetic counterparts can be used to measure the relative speeds, but only if the intrinsic time-lag between emission of the photons and gravitational waves is well understood. Here we suggest a method that does not make such an assumption, using future strongly lensed GW events and EM counterparts; \citet{biesiada} forecast that 50-100 strongly lensed GW events will be observed each year with the Einstein Telescope. A single strongly lensed GW event would produce robust constraints on $c_{\mathrm{GW}}/c_{\gamma}$ at the  $10^{-7}$ level, if a high energy EM counterpart is observed within the field-of-view of an observing gamma ray burst monitor.
\end{abstract}

\setcounter{footnote}{1}
\maketitle

\section{I. Introduction}
\label{sec:intro}

On September 14, 2015 at 09:50:45 UTC the two detectors of the Laser Interferometer Gravitational-Wave Observatory observed a transient gravitational-wave signal from a black hole--black hole binary (BHBH) inspiral \citep{ligo}, opening the era of gravitational wave astronomy. Despite extensive searches for electromagnetic counterparts \citep{followup} none have been conclusively detected (although a possibly gamma ray counterpart was identified in \citep{fermi}). The presence of an optical counterpart to a BHBH would be somewhat surprising, although models have been suggested by \citet{loeb,perna} and \citet{janiuk} that can generate EM counterparts to stellar mass BHBH mergers. However gravitational wave (GW) events associated with double compact object inspirals, are expected to produce large amounts of Electromagnetic radiation \citep{sylvestre}.

The initial GW detection has already enabled a suite of tests of general relativity \citep{ligograv}. However, if EM and GW signals are detected from the same transient phenomenon, this opens up a new physical window with which to test the relative speeds of light and gravitational waves. The consistency of the LIGO GW event and event templates predicted by general relativity places tight constraints on the nature of gravity \citep{ligograv} in the strong field limit. Several of the alternative theories of gravity invoked to explain the accelerated expansion of the Universe \citep{riess,perlmutter} require deviations in the weak field limit, and many of these theories predict $c_{\mathrm{GW}} \neq c_{\gamma}$ \citep{gleyzes,felice,bellini}. The absence of detectable dispersion in the LIGO GW signal places a tight limit on graviton mass \citep{ligograv}, but does not test the propagation speed of GWs relative to photons. Constraining the parameter 
\be
\symbl = \frac{c_{\mathrm{GW}}}{c_{\gamma}},
\ee
is therefore an important test of the modified gravity theories that may mimic dark energy.

The absence of gravitational Cherenkov radiation from cosmic rays demonstrates that GW cannot travel significantly slower than the speed of light \citep{caves}; \citet{moorenelson} showed that $1- c_{\mathrm GW}/c_{\gamma} < 10^{-15}$. \citet{beltran} also recently used observations of the Hulse-Taylor pulsar to constrain the gravitational wave speed to be greater than 0.995 $c_\gamma$. \citet{blas} have used the 6.9 ms time delay between the GW signals detected at the two LIGO observatories to show that  $c_{\mathrm{GW}}/c_{\gamma}<1.7$, but this loose bound can be improved upon significantly if we can measure the difference in travel times for photons and GWs over cosmological baselines. In this case $\symbl$ is given by
\be
1-\symbl^{-1}= c_{\gamma} \Delta t / D_{\mathrm{proper}},
\ee
where $\Delta t$ is the difference in travel times between GW and photons and $D_{\mathrm{proper}}$ is the proper distance to the source. However $\Delta t$ is only the difference in arrival times if the EM and GW waves are emitted simultaneously and at the same location. Otherwise any difference in the arrival time of the EM and GW components may be due to $\symbl \neq 1$ or an emission lag in the source. 

In this paper we instead propose a novel measurement exploiting strongly lensed GW events to measure $\symbl$, requiring no knowledge of the emission mechanism. It is reasonable to consider such a measurement as strongly lensed GW events are likely to be observed with the next generation of GW detectors \citep{biesiada}.

\section{II. A test of ${c_{\mathrm GW}}/{c_{\gamma}}$ with a single strongly lensed source}

A measurement of ${c_{\mathrm GW}}/{c_{\gamma}}$ is possible if a GW source and EM counterpart are multiply imaged by a strong gravitational lens. In this case, robust constraints can be made without strong assumptions about the connection between the EM and GW emission.

The light travel time through an inhomogeneous Universe is a function of both geometrical distance and Shapiro delay. Rays travelling along different paths take different times to reach the same observer. According to Fermat's principle, images only form at extrema of this time-delay function \citep{blandford}, so for most transient events only one image is observed. When an event occurs behind a strong gravitational lens, the time-delay surface is sufficiently distorted that multiple extrema, and hence multiple images, form. Time-variable events are not observed simultaneously in each image due to the different paths that the rays have travelled. This extra travel time allows a test of $c_{\mathrm{GW}}$, limited by our ability to measure the difference between the EM and GW time-delays.

For a strong gravitational lens, general relativity makes three relevant predictions:
\begin{itemize}
{\item  {light and GWs travel at the same speed,}}
{\item  {light and GWs travel along the same paths, and}}
{\item  {light and GWs feel the same Shapiro delay.}}
\end{itemize}

Suppose there is a lens where two images are observed, and the light travel time for a photon along each path is $\tau$ and $\tau+\Delta\tau$. Further assume an EM and GW event occur in the source with unknown initial time separation $\Delta t_{\mathrm{int}}$. 
Then if GR is correct, the observer will see an EM event at $t_0$, a GW event at $t_1=t_0 + (1+z_s)\Delta t_{\mathrm{int}}$, a second EM event at $t_2=t_0+\Delta\tau$ and a second GW event at $t_3=t_0 + (1+z_s)\Delta t_{\mathrm{int}} + \Delta\tau$. The first three events uniquely solve for $t_0$, $(1+z_s)\Delta t_{\mathrm{int}}$ and $\Delta\tau$, so the second GW event is a robust test of GR.

If any of the GR predictions are violated, the second GW event will not occur at $t_0 + (1+z_s)\Delta t_{\mathrm{int}} + \Delta\tau$. To interpret any difference however requires us to make assumptions about which of the predictions are broken. 

Many modified gravity theories \citep{modg1} invoke screening mechanisms that predict deviations from  $\symbl=1$ \citep{modg2}. Assuming that any observed difference in the GW and EM time-delays is entirely due to $c_{\mathrm GW} \neq c_{\gamma}$, we find that the ratio of the speeds is given by
\be
\frac{c_{\mathrm{GW}}}{c_{\gamma}} = \frac{t_2-t_0}{t_3-t_1}.
\ee
This probe of $\symbl$ is only a function of the time that the signals are observed, and hence does not require knowledge of the position of the inspiral on the sky.

Massive gravity theories \citep{MassiveG} potentially allow for violations of all three of the GR predictions listed above, and converting differences in observed time delays into constraints on these theories requires careful attention. Directly testing the assumption that the rays travel along the same paths is practically impossible, since typical image separations in a strong lens are of order arc seconds, and such angular resolution is impossible with the GW detectors currently built/planned. One might hope that observing the flux ratios of the EM and GW events might test if the rays travelled similar paths. Unfortunately microlensing by stars in the lens galaxy will cause variation in flux-ratios unless the EM and GW emission regions are in exactly the same location and are the same size \citep[see][for a demonstration of this kind of chromatic microlensing]{blackburne}. 

Whilst observing the exact location of GW images is beyond current technology, it is possible to predict their locations if the mass distribution of the lens and the unlensed source position are known. In this case, one can calculate the deflection angles and Shapiro delays experienced by photons and gravitons as they pass through the lens even for modified gravity theories that break all of the GR predictions listed above.  With known EM image locations, the lens equation for photons can be used to solve for the location of the optical component in the source plane. Assuming the GW emission comes from the same location on the source plane as an EM counterpart, it is then possible to solve for the location at which the GW images must have formed using the modified-gravity lens equation for the gravitons. Doing this precisely requires well determined EM image locations (i.e. an afterglow must be found), and knowledge of the lens matter distribution. Precisely inferring the mass distribution of the lens is possible with high resolution imaging of an extended EM component \citep[such as a host galaxy, e.g.][]{suyu2014,collett2014}. Observed constraints on ${t_2-t_0}={t_3-t_1}$ can therefore be mapped into constraints on the massive gravity theories. However, significantly more work and ancillary data are needed for testing this class of theories, than are needed for probing $\symbl$ in the context of gravity theories where the EM and GW follow the same path and feel the same Shapiro delay.

\section{III. Plausibility}
A single strong lensing event with time delays measured in both EM and GW is guaranteed to give good constraints on ${c_{\mathrm{GW}}}/{c_{\gamma}}$. As we noted above, the Einstein telescope, which is a plausible evolution of the LIGO/VIRGO concept, is forecast to discover 50-100 strongly lensed GW events per year \citep{biesiada}. In light of updated estimates of GW rates \citep{belczynski}, this forecast is likely to be an underestimate. In order to calculate the constraint on $\symbl$, we note that typical time-delays for strong lensing by galaxies and clusters are of order hours to months \citep{OM10}, while  the chirp of a GW event can be measured at millisecond precision \citep{ligo}. Extremely precise constraints on  ${c_{\mathrm{GW}}}/{c_{\gamma}}$ are therefore possible, if the precision of the EM time-delay approaches that of the GW time-delays. EM counterparts are expected to be ubiquitous for several classes of double compact object inspiral \citep{sylvestre}, and an immediate optical follow-up campaign such as \citet{des} could be expected to measure optical time-delay determinations accurate to a few hours depending on the variability of the EM transient \citepeg{bonvin}. The most exquisite constraints are however possible if the EM event occurs within the field-of-view of an already observing telescope that has good temporal resolution. Gamma ray burst monitors have large fields of view and are capable of timing events with sub-second precision, although it is necessary that both EM signals are observed at this high precision - this is potentially a challenge if the field-of-view of the telescope changes over the duration of the strong lensing time delay. For a strongly lensed GW event with a one month time-delay, measured with 0.1 second precision on the EM delay and millisecond precision in the GW delay, we would obtain a constraint on ${c_{\mathrm{GW}}}/{c_{\gamma}}$ with uncertainty
\be
\sigma\left(\frac{c_{\mathrm{GW}}}{c_{\gamma}}\right) \approx 10^{-7}.
\ee
Such a measurement therefore provides a stringent test of whether light and gravitational waves travel at the same speed, and can be confidently used to place constraints on any theory of gravity. Since the method relies only on the time separation of the events, it does not require any knowledge of the sky position of the inspiral event. 

\section{IV. Identifying lensed GW events and their EM counterparts}

One challenge for measuring $\symbl$ using strongly lensed GWs is identifying the events. The era of the Einstein telescope will potentially see thousands of events per year - identifying 2 lensed images of the same event separated by a month will therefore pose a statistical challenge. A network of two (four) 3rd generation GW detectors will typically give positional accuracy to 100 (10) square degrees for sources out to redshift 3 (6)  \citep{vitaleevans2016} so less than 0.2 (0.02) percent of GW events will be from mutually consistent sky locations. The cross matching is made easier since lensing is achromatic: the observed strains must be the same up to their amplitude. Identifying a lensed GW event should not therefore pose great difficulty. However it is worth noting that chirps will not have the same recovered physical parameters.: magnified chirps look the same as unmagnified chirps originating from more massive events at lower redshifts \citep{dai2016}.

If $\symbl$ is close to 1, the approach for identifying EM counterparts is the same as for identifying EM counterparts for an unlensed GW event. The task is made slightly easier since the flux ratios and time delays will be approximately the same for the EM and GW events.

If $\symbl$  deviates significantly from 1, cross matching EM events with GW events will be extremely difficult unless the EM sources associated with GW emission are already known and differ significantly from other EM sources: other lensed sources are vastly more common \citep{lenspop, OM10} and for short time delays every galaxy in the universe could potentially be the lens. However, only the most massive objects can cause long time delays \citep{OM10} and these are extremely rare, so the lensed GW events that have the potential to give the best constraints on $\symbl$ are also the easiest lenses to identify.

\section{V. Conclusion}

In this letter we have proposed a method for robustly measuring the speed of  gravitational waves, with no need to understand the EM emission mechanism. While we have framed this letter in the context of testing if ${c_{\mathrm{GW}}}={c_{\gamma}}$, the argument presented is completely generic for testing relative speeds of any waves or particles without making strong assumptions about the emission lag. One example that is certainly plausible with current technology would be to test for Lorentz equivalence violations by measuring if the speed of light varies with photon energy using a strongly lensed gamma ray burst \citep{biesiada2}.

One further technical barrier to the measurement of $\symbl$ is the need for a suitable all sky, high temporal resolution EM detector. The field-of-view is especially important, since both images of the EM event must be detected, potentially months apart. If no wide field gamma ray monitors are available, then the constraints derivable from optical followup alone will only be able to measure $\symbl$ at the $\approx 10^{-2}$ level.

The method does however make the assumption that the EM and GW are generated at similar locations. For a typical lens with a 1 month time delay \footnote{Singular isothermal sphere with velocity dispersion of 220 km s$^{-1}$, $z_l=0.4$, $z_s=1$, unlensed source position offset from the lens centre by half the Einstein radius}, a spatial offset of 60 AU in projection between the EM emitter and the inspiral event would also cause a relative difference in the time delays of $10^{-7}$. 60 AU is much larger than the scales relevant to stellar mass inspiral events, but this uncertainty will place a floor on the precision of ${c_{\mathrm{GW}}}/{c_{\gamma}}$ constraints that are derivable from differential time-delays even with perfect time resolution of both the GW and EM events.

If gamma ray counterparts are commonly associated with gravitational wave emission, it is only a matter of time before a strongly lensed event is detected. Robust and precise constraints on ${c_{\mathrm{GW}}}/{c_{\gamma}}$ will therefore be achieved if associated EM emission is also detected. 


\section*{Acknowledgements}
\noindent  We thank the referees for their useful comments that have improved this manuscript. We have enjoyed fruitful conversations about this work with David Wands, Jeremy Sakstein, Bob Nichol, John Ellis and Shantanu Desai. We are grateful to them. We thank Chuck Keeton and Sean Brennan for prompting us to investigate how differential source locations can produce differential time delays.
TC acknowledges funding for this work from the University of Portsmouth. DB is supported by STFC Consolidated Grant ST/N000668/1.








\label{lastpage}

\end{document}